\documentclass[aps,prl,superscriptaddress,twocolumn,showpacs,amsmath,floatfix]{revtex4-2}

\usepackage{float}
\usepackage{tikz-cd}
\usepackage{tikz}
\usepackage{adjustbox}
\usetikzlibrary{graphs,decorations.markings,calc,backgrounds}
\usepackage{amssymb}
\usepackage{etoolbox}

\usepackage{graphicx}
\usepackage{amsmath,amsthm,amsfonts,amssymb} 
\usepackage[normalem]{ulem} 
\usepackage{bm} 
\usepackage{multirow} 
\usepackage{color} 
\usepackage{makeidx} 
\usepackage[linewidth=1pt,nobreak=true]{mdframed} 
\usepackage[titletoc]{appendix} 
\usepackage{enumitem} 
\usepackage{tikz-cd}  
\usepackage[colorlinks]{hyperref} 

\allowdisplaybreaks[4]  

\begin{document}

\title{Beyond Boundaries: efficient  Projected Entangled Pair States methods for periodic quantum systems}

\author{Shaojun Dong}\thanks{The two authors made equal contributions.}
\affiliation{Institute of Artificial Intelligence, Hefei Comprehensive National Science Center, Hefei, 230088, People's Republic of China}

\author{Chao Wang}\thanks{The two authors made equal contributions.}
\affiliation{Institute of Artificial Intelligence, Hefei Comprehensive National Science Center, Hefei, 230088, People's Republic of China}

\author{Hao Zhang}
\affiliation{CAS Key Laboratory of Quantum Information, University of Science and Technology of China, Hefei 230026, People's Republic of China}
\affiliation{Synergetic Innovation Center of Quantum Information and Quantum Physics, University of Science and Technology of China, Hefei 230026, China}

\author{Meng Zhang}
\affiliation{Institute of Artificial Intelligence, Hefei Comprehensive National Science Center, Hefei, 230088, People's Republic of China}

\author{Lixin He}
\email{helx@ustc.edu.cn}
\affiliation{CAS Key Laboratory of Quantum Information, University of Science and Technology of China, Hefei 230026, People's Republic of China}
\affiliation{Institute of Artificial Intelligence, Hefei Comprehensive National Science Center, Hefei, 230088, People's Republic of China}
\affiliation{Synergetic Innovation Center of Quantum Information and Quantum Physics, University of Science and Technology of China, Hefei 230026, China}
\affiliation{Hefei National Laboratory, University of Science and Technology of China, Hefei 230088, China}

\begin{abstract}

Projected Entangled Pair States (PEPS) are recognized as a potent tool for exploring two-dimensional quantum many-body systems.
However, a significant challenge emerges when applying conventional PEPS methodologies to systems with periodic boundary conditions (PBC),
attributed to the prohibitive computational scaling with the bond dimension. This has notably restricted the study of systems with complex boundary conditions.
To address this challenge, we have developed a strategy that involves the superposition of PEPS with open boundary conditions (OBC) to treat systems with PBC. This approach significantly reduces the computational complexity of such systems
while maintaining their translational invariance and the PBC.
We benchmark this method against the Heisenberg model and the $J_1$-$J_2$ model, demonstrating its capability to yield highly accurate results at low computational costs, even for large system sizes.
The techniques are adaptable to other boundary conditions, including cylindrical and twisted boundary conditions, and therefore significantly expands the application scope of the PEPS approach, shining new light on numerous applications.
\end{abstract}

\maketitle

{\it Introduction:} Interacting quantum many-body systems are at the forefront of some of the most intriguing and challenging problems in contemporary physics. The phenomena arising from quantum many-body effects in two-dimensional (2D) condensed matter are central to a multitude of groundbreaking discoveries in the field. These include the exploration of quantum spin liquids \cite{savary2016quantum}, the elucidation of high-Tc superconductivity \cite{lee2006doping}, the investigation of the fractional quantum Hall effect \cite{stormer1999fractional}, and the study of string-net condensation \cite{levin2005string,gu2009tensor}. Simulating strongly correlated systems is not only crucial for understanding these complex phenomena but also poses one of the most significant challenges in condensed matter physics.

The projected entangled pair states (PEPS) methods \cite{Orus14,Vidal03,Verstraete04,Vidal08,Xie14,Schollwoeck2011,Garcia2006,Verstraete2008,Xiang08,Sfondrini10,Verstraete06}  have shown their power on simulation of the strongly correlated many-particle systems on 2D lattice\cite{Liu18,Liao17,Coboz14,Dong20} ,
which have achieved great success.
However, so far the PEPS methods are mostly widely applied to systems with infinite sizes \cite{Orus08,Xiang08,Coboz10,Orus15,Coboz16},
or system with the open boundary conditions (OBC) \cite{Cirac14,liu2017gradient,liu2022gapless}, attributed to the prohibitive computational scaling with the bond dimension
for periodic boundary conditions (PBC) .

Simulations of finite lattices with PBC are crucial for analyzing various quantum systems.
A common approach to approximating physical properties at the thermodynamic limit involves the finite-size scaling of these systems.
Unlike OBC, which suffer from significant boundary effects,  PBC exhibits much weaker boundary influences,
allowing for faster convergence with increasing system size.
More importantly, the investigation of topological order in quantum systems requires a non-trivial topological base manifold to facilitate topological degeneracy. This requirement is essential for the of topological invariants, such as the $U$ and $S$ matrices, which are vital for classifying topological orders \cite{moradi2015}.  In this context, PBC offer an ideal framework for analyzing topological phenomena.
In the study of topological order, such as fractional quantum Hall effect (FQHE), computing many-body Chern numbers often requires the use of twisted boundary conditions \cite{niu1985}, which are modifications of PBC. The versatility of PBC in accommodating various boundary scenarios significantly enhances its utility in the exploration of complex quantum phenomena. Consequently, PBC plays an indispensable role in these studies.

\begin{figure}[t!]
	\centering
	\includegraphics{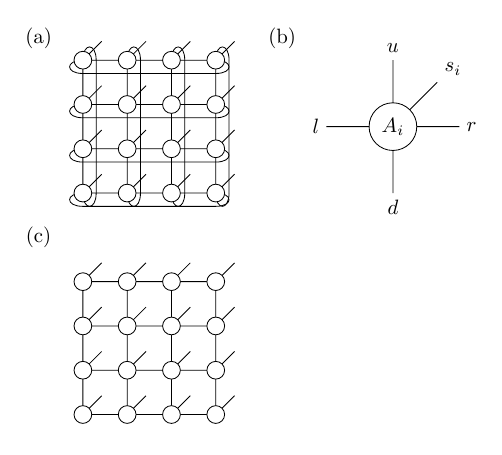}
	\caption{(a) Illustration of a naive PEPS with PBC (pPEPS) on a square lattice. (b) A single tensor at position $\bm{r}_i$, with four virtual bonds $l$, $r$, $u$, $d$ of dimension $D$, and a physical index $s_i$. (c) Illustration of a PEPS with OBC (oPEPS). The physical interactions can be either OBC or PBC. }
	\label{fig:peps}
\end{figure}

Although PBC is crucial in the study of quantum many-body problems, implementing efficient PBC algorithms in PEPS is an extremely challenging task.
A naive method for constructing a PEPS for systems with PBC is illustrated in Fig.~\ref{fig:peps}(a).
This method adheres to the area law of entanglement entropy \cite{eisert2010colloquium} and preserves translational symmetry. We refer to this PEPS construction as periodic PEPS (pPEPS).
However, directly contracting such a tensor network incurs exorbitant computational costs, scaling at $\sim O(D^{18})$ \cite{Sfondrini10},  where $D$ is the virtual bond dimension.
Although employing Monte Carlo (MC) sampling techniques\cite{sandvik2007variational,liu2017gradient} can significantly reduce this scaling, the computational demands remain prohibitively high for practical analysis of significant physical problems. The application of PEPS in cylindrical boundary conditions, which are periodic along one direction, also encounters this problem \cite{Cirac11,Coboz17}. Due to these computational complexity issues, PEPS with PBC has not yet been widely adopted, significantly limiting the applicability of the PEPS method in the study of quantum many-body problems. This has become a critical challenge that urgently needs to be addressed in the field.

In this letter, we introduce an ansatz that extends PEPS, specifically designed for translationally invariant lattice models with PBC. Our ground state optimization algorithm, which utilizes this ansatz, achieves computational costs of $O(D^6)$, significantly lower than those associated with the naive implementation of pPEPS. This advancement overcomes a significant challenge in the PEPS methodology, enabling efficient simulation of periodic systems.  Consequently, it greatly expands the applicability of the PEPS approach, facilitating the exploration of novel physical phenomena, including topological orders, using PEPS.

{\it Methods:}
Consider a many-body model defined on a two-dimensional (2D) lattice with PBC,  which possesses translational symmetry. We define a supercell of dimensions $L_1 \times L_2$, with PBC.  Let $|{s_i}\rangle$ denotes the physical state at lattice site $i$.
For a spin $1/2$ system, $s_i$= $\uparrow$, $\downarrow$ represents the spin states at  $\bm{r}_i$.

Direct application of  pPEPS to study this model would suffer from extremely high computational complexity,
making this approach impractical.
 As a compromise, one can use PEPS with open boundary conditions (oPEPS),
 as depicted in Fig.~\ref{fig:peps}(c), to approximate the wave function while preserving periodic physical interactions.
The wave function reads,
\begin{equation}
  |\Psi_{\rm oPEPS}\rangle = \sum_{s} { \rm Tr} \big(\prod_i A_{i}^{s_{i}} \big) |s \rangle,
   \label{Eq:PEPS}
\end{equation}
where $|s\rangle=|s_1,s_2,\cdots\rangle$ is the spin configuration, and $A^{s_{i}}_{i}$=$(A_{i})_{l,r,u,d,s_{i}}$ is a five-index tensor located on site $i$ as shown in Fig.~\ref{fig:peps}(b).
It has a physical index $s_{i}$ of dimension $d_p$ and four virtual indices $l, r, u, d$, each of dimension $D$, corresponding to the four nearest neighbors.

We will demonstrate that for small systems, it is still feasible to obtain high-quality ground states in a periodic system using oPEPS, provided that the bond dimension  $D$ is sufficiently large.
However, this approach becomes less effective as the system size increases for several reasons. Firstly, oPEPS lacks translational symmetry. Moreover, for large systems (supercells), the required $D$ for oPEPS to converge increases rapidly. This is due to the need to capture correlations between distant boundaries through a long-range network, which suggests that this method is unsuitable for periodic systems.

\begin{figure}[t!]
	\centering
	\includegraphics{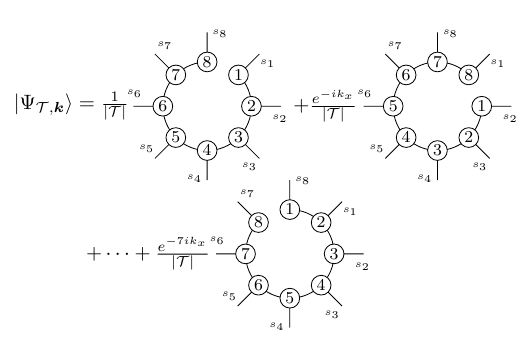}
	\caption{A superposed PEPS (sPEPS) on a $1 \times 8$ lattice, where the numbers in the circles represent the tensors, and $s_i$ are the spin indices. The sPEPS is obtained by applying the Fourier transformation of the shifting operators to an oPEPS.
}
	\label{fig:shifted_peps}
\end{figure}

Now, we shall introduce a new form of PEPS that can preserve translational symmetry and satisfy the PBC. We define a translation operator,
\begin{align}
\hat{T}_{j} &= \sum_{s'}\sum_{s}\Big[\prod_{i}\delta_{s_{i-j},s'_{i}}\Big]|s'\rangle\langle s|
\end{align}
which translates the total wave function of the quantum system by $\bm{r}_j$.

We apply $\hat T_{j}$ to the oPEPS, which gives,
\begin{align}
	\hat T_{j}|\Psi_{\rm oPEPS}\rangle = &\sum_{s} { \rm Tr}\big(\prod_k A_{k}^{s_{i+j}} \big)|s\rangle  \nonumber\\
	=&\sum_{s} { \rm Tr}\big(\prod_i A_{i-j}^{s_{i}} \big)|s \rangle ,
\end{align}
i.e., it translates the spin indices by $j$ or equivalently translates the tensors' indices by $-j$.
It is easy to show that the translation operators $\hat T_j$
are unitary and form the translational group $\mathcal T$ of the system that commutes with the Hamiltonian $H$.

We can define a many-body wave function that satisfies PBC using the translation operators,
\begin{equation}
\label{eq::sPEPS}
    |\Psi_{\mathcal{T}, \bm{k}}\rangle = \hat{T}_{\mathcal{T}, \bm{k}} |\Psi_{\rm oPEPS}\rangle,
\end{equation}
where $\hat{T}_{\mathcal{T}, \bm{k}}$ is the Fourier transform of the translation operators with momentum $\bm{k}$,
\begin{equation}
\label{eq::shifting}
\hat{T}_{\mathcal{T}, \bm{k}} = \frac{1}{|\mathcal{T}|} \sum_{i} e^{-i \bm{k} \cdot \bm{r}_i} \hat{T}_{i},
\end{equation}
and $i$ runs over the entire lattice of the supercell. $|\mathcal{T}|$ is the number of translation symmetry operations.
The values of $\bm{k}$ satisfy the von Neumann boundary conditions.
We name the wave function of Eq. (\ref{eq::sPEPS}) the superposed PEPS (sPEPS) with momentum ${\bm k}$.
We illustrate the sPEPS using an example $1\times8$ system in Fig.~\ref{fig:shifted_peps}.
It is easy to shown that the sPEPS ansatz respects the translational symmetry of the system.

As discussed in Ref.~\cite{Supplementary}, $\hat{T}_{\mathcal{T}, \bm{k}}$ is a projector into the subspace with a definite momentum $\bm{k}$.
The ground state of a translationally invariant Hamiltonian $H$ possesses a certain momentum $\bm{k}$, which may be nonzero in the presence of a gauge field
or spin-orbit coupling. Applying $\hat{T}_{\mathcal{T}, \bm{k}}$ to a tensor network state ensures that the resulting state has the desired momentum.

 In practice, to reduce computational time, we may select a subgroup
 $\mathcal{S} \subseteq \mathcal{T}$
  instead of the full translation group $\mathcal{T}$ to construct sPEPS. Specifically,
\begin{equation}
\label{eq:proj_subg}
|\Psi_{\mathcal{S}, \bm{k}}\rangle = \hat{T}_{\mathcal{S}, \bm{k}}|\Psi_{\rm oPEPS}\rangle = \frac{1}{|\mathcal{S}|} \sum_{j \in \mathcal{S}} e^{-i \bm{k} \cdot \bm{r}_j} \hat{T}_{j} |\Psi_{\rm oPEPS}\rangle,
\end{equation}
where $|\mathcal{S}|$ is the size of $\mathcal{S}$.
 A smaller subgroup $\mathcal{S}$ reduces computational cost but may lower accuracy.
Therefore, choosing a suitable subgroup $\mathcal{S}$ is a balance between accuracy and efficiency.
After the simulation process, we can further improve accuracy by using the full projection\cite{Supplementary}:
\begin{equation}
 |\Psi_{\mathcal{T},\bm{k}}\rangle = \hat{T}_{\mathcal{T},\bm{k}} |\Psi_{\mathcal{S},\bm{k}}\rangle
 = \hat{T}_{\mathcal{T},\bm{k}} |\Psi_{\rm{oPEPS}}\rangle.
\label{eq:full_proj}
\end{equation}
We note that the wave function $|\Psi_{\mathcal{T},\bm{k}}\rangle$ obtained in Eq.~(\ref{eq::sPEPS}) may not be equivalent to that from Eq.~(\ref{eq:full_proj}), as $|\Psi_{\rm{oPEPS}}\rangle$ could differ due to the different optimization processes.

To obtain the ground state, we first perform the imaginary time evolution method with the simple update (SU) \cite{Xiang08} algorithm, which is a fast local optimization method.
For a translationally invariant Hamiltonian $H$ with  PBC, the imaginary time evolution operator commutes with the translation operators $\hat{T}_{i}$. Therefore, we have:
\begin{equation}\label{eq::SU}
e^{-\tau H}|\Psi_{\mathcal{S},\bm{k}}\rangle = \hat{T}_{\mathcal{S},\bm{k}} \left( e^{-\tau H}|\Psi_{\rm oPEPS}\rangle \right).
\end{equation}
This means the SU on the sPEPS can be performed by first applying the SU to $|\Psi_{\rm oPEPS}\rangle$, followed by the action of the translation operator in Eq.~(\ref{eq:proj_subg}). Consequently, the computational cost of the SU in sPEPS is the same as that of the oPEPS. However, during SU, the correlations between boundary sites are treated as long-range correlations via the oPEPS and are not handled very well.

The SU optimization gives us an approximation of $|\Psi_{\mathcal{S},\bm{k}}\rangle$.
We can quickly screen the ground-state momentum $\bm{k}$ through these results.
We then perform the gradient optimization (GO) process for further refinement. To reduce the high scaling of the contraction, Monte Carlo (MC) sampling techniques are used, similar to those in oPEPS\cite{liu2017gradient}. The only difference is that the wave function is replaced by $|\Psi_{\mathcal{S},\bm{k}}\rangle$.
 The details are explained in \cite{Supplementary}.
The computational cost remains $O(D^6)$ for a given $D$, where $D$ is the bond dimension.
Once we obtain the optimal $|\Psi_{\mathcal{S},\bm{k}}\rangle$, we perform a full projection to obtain the ground state sPEPS $|\Psi_{\mathcal{T},\bm{k}}\rangle$ with better accuracy, using Eq. (\ref{eq:full_proj}).

{\it Numerical Validation---}We benchmark our sPEPS algorithm on the spin-1/2 $J_1$-$J_2$ model with PBC.
The Hamiltonian of the model reads,
\begin{equation}
    H=J_1\sum_{\langle i, j\rangle}{\bm s}_i\cdot {\bm s}_j+J_2\sum_{\langle \langle i, j\rangle\rangle}{\bm s}_i\cdot {\bm s}_j ,
\end{equation}
where $\langle i, j\rangle$ stands for nearest neighbors, $\langle \langle i, j\rangle\rangle$ stands for next-nearest neighbors, and the ${\bm s}_i$ are the spin operators defined at site $i$. In the simulations, we fix $J_1$=1.

The $J_1$-$J_2$ model with OBC has been studied using oPEPS \cite{liu2022gapless,Liu18,Dong20}.
In this work, we simulate the ground state of the $J_1$-$J_2$ model with PBC on an $L \times L$ square lattice using sPEPS.
To reduce computational costs, we use a subgroup of the full translation symmetry. We shift the oPEPS along the diagonal direction, which yields satisfactory results. The computational cost increases linearly with system size $L$, instead of $L^2$.

We first examine the model at $J_2 = 0$, where it simplifies to the antiferromagnetic Heisenberg model.
We benchmark the ground state energy using a bond dimension $D = 6$ for various system sizes, employing both the oPEPS and sPEPS ansatz.
We identify that the ground states have a momentum $\bm{k}$=0 for the sPEPS.
We compare our results with the QMC results\cite{sandvik1997finite}, which have an accuracy   $\sim 10^{-4}$ for even a small $D=6$. The results are listed in Table~\ref{tab:Heisenberg}. Both ansatzes perform well in the $4 \times 4$ system, where the error is less than $10^{-4}$ compared with QMC.
However, the accuracy of the oPEPS ansatz diminishes as the system size increases. In the largest system of $14 \times 14$, the accuracy of the oPEPS ansatz degrades to $5 \times 10^{-3}$ because it cannot adequately capture the correlations in the periodic system. On the other hand, the sPEPS ansatz maintains consistent accuracy with increasing system sizes, achieving an accuracy of $10^{-4}$ on the $14 \times 14$ lattice.

\begin{table} [tb!]
\caption{Comparison of ground-state energies of the Heisenberg model calculated by oPEPS, sPEPS, and QMC simulations on $L \times L$ square lattices, with $L$ ranging from 4 to 14. The bond dimension is set to $D=6$ in PEPS.}
\begin{tabular}{ c c c c}
\hline\hline
$L$  & oPEPS & sPEPS & QMC\cite{sandvik1997finite}  \\
\hline
4   & -0.70170 & -0.70176 & -0.70178 \\
6   & -0.67854 & -0.67885 & -0.67887 \\
8   & -0.67127 & -0.67335 & -0.67348 \\
10  & -0.66985 & -0.67143 & -0.67155 \\
12  & -0.66812 & -0.67056 & -0.67069 \\
14  & -0.66679 & -0.67000 & -0.67022 \\
\hline
\hline
\end{tabular}
\label{tab:Heisenberg}
 \end{table}

We then study the model at $J_2 = 0.5J_1$, with $L = 6$ and $L = 10$.
In this parameter region, the model is reported to be in a quantum spin liquid phase \cite{Wang18, Ferrari20, liu2022gapless}.
Due to the strongly frustrated interactions and correlations, this case is numerically challenging for many widely used numerical methods. In particular, QMC fails because of the notorious sign problem \cite{Loh1990, Troyer05}.

We list the ground state energies obtained by sPEPS at $J_2 = 0.5J_1$ in Table~\ref{tab:J1J2}, comparing them with results from the density matrix renormalization group (DMRG) \cite{gong2014plaquette}, convolutional neural network (CNN) \cite{choo2019two,liang2021hybrid,Liang_2023}, and exact diagonalization (ED) methods \cite{schulz1996magnetic}, etc.
For the sPEPS calculations, we use a modest bond dimension $D$=8, and the ground states have a momentum $\bm{k}$=0.

 For the $6 \times 6$ system, the exact value of the ground state energy, $E = -0.50381$, is obtainable from ED. The energy obtained from sPEPS is $E = -0.50375$, which is slightly higher than the DMRG result of $E = -0.50380$ \cite{gong2014plaquette} and much better than the CNN result of $E = -0.50185$ from Refs. \cite{choo2019two}.

For the $10 \times 10$ system, the exact ground state energy is not available. The energy obtained from sPEPS is $E = -0.49759$, which is significantly better than the DMRG result of $E = -0.49553$ \cite{gong2014plaquette}. It is well-known that DMRG performs effectively for narrow systems; however, its accuracy diminishes rapidly as the system width increases\cite{Schollwoeck2011}. To maintain accuracy, the number of kept states must grow exponentially with the system's width. Our result is also significantly superior to the earlier CNN results \cite{choo2019two,liang2021hybrid}.

The Lanczos procedure can significantly improve the energy, as demonstrated in Refs. \cite{hu2013direct} for the variational Monte Carlo (VMC) method and in Ref. \cite{Liang_2023} for the CNN method. However, our results outperform these approaches and are also comparable to the state-of-the-art ground state energy $E = -0.497629$ achieved by the PP$+$RBM method \cite{RBM}.

We note that CNN and other neural network methods are powerful and flexible methods for studying quantum many-body problems and has made significant progress in recent years\cite{choo2019two,liang2021hybrid,Liang_2023}.
However, due to its difficulty in handling the spin sign structure\cite{Westerhout2020}, CNN and PEPS can be considered complementary methods, can be used together to provide a more comprehensive understanding of quantum many-body systems \cite{liang2021hybrid}.

\begin{table} [tb!]
\caption{Comparison of ground-state energies of the $J_1$-$J_2$ model with $J_2 = 0.5J_1$ on a $10 \times 10$ lattice, obtained from sPEPS ($D=8$), DMRG, and various Neural Network methods.}
\begin{tabular}{ c c c }
\hline\hline
method & Energy(L=6)  & Energy(L=10)\\
\hline
sPEPS  & -0.50375 & -0.49759 \\
DMRG\cite{gong2014plaquette} &-0.50380 & -0.49553 \\
CNN\cite{choo2019two} &-0.50185  & -0.49516 \\
CNN\cite{liang2021hybrid} & - & -0.49550 \\
CNN\cite{Liang_2023}  & -& -0.49747 \\
VMC\cite{hu2013direct} & - & -0.49755 \\
PP+RBM\cite{RBM}  & -& -0.49763 \\
exact\cite{schulz1996magnetic} & -0.50381 &- \\
\hline
\hline
\end{tabular}
\label{tab:J1J2}
 \end{table}

\begin{figure}[hbt!]
  \includegraphics[width=0.49\textwidth]{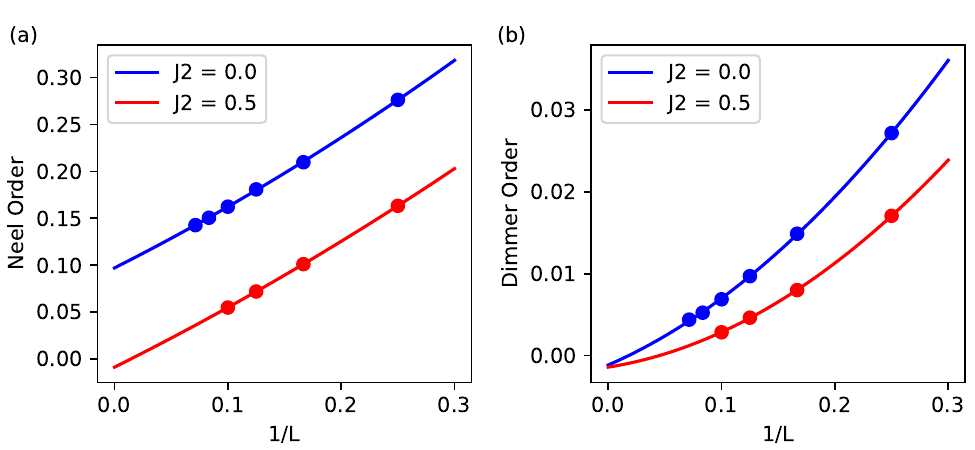}
 \caption{(a) The N\'{e}el order parameters and (b) the dimer order parameters for $J_2 = 0$ (blue) and $J_2 = 0.5J_1$ (red) as functions of $1/L$.}
  \label{fig:order_para}
\end{figure}

We further calculate the spin structure factor $m_{\bm{k}}$ at momentum ${\bm{k}}$, which is defined by
\begin{equation}
    m_{\bm{k}}^2 = \frac{1}{L^4} \sum_{i,j} e^{2\pi i({\bm r}_i-{\bm r}_j)\cdot{\bm k} } \langle {\bm{s}}_{i} \cdot {\bm{s}}_{j} \rangle.
\end{equation}
In particular, for ${\bm{k}} = (\pi, \pi)/L$, $m_{\bm{k}}$ is the N\'{e}el order parameter, which serves as the order parameter for the antiferromagnetic (AFM) phase. We calculate the N\'{e}el order parameter at $J_2 = 0$ and $J_2 = 0.5J_1$ for various system sizes. Quadratic finite-size scaling is performed to extrapolate the results to the thermodynamic limit ($L \rightarrow \infty$).
The results are plotted in Fig.~\ref{fig:order_para}(a). In the thermodynamic limit, the N\'{e}el order parameter is $m_s = 0.311$ for $J_2 = 0$ and $m_s = 0$ for $J_2 = 0.5$, which is consistent with reports that the model is in an AFM phase at $J_2 = 0$ and lacks spin order at $J_2 = 0.5J_1$. \cite{liu2022gapless}

We also calculate the dimer order parameter $D^\alpha_{\bm{k}}$ ($\alpha=x/y$) at momentum ${\bm{k}}$, which is defined by
\begin{equation}
	D^\alpha_{\bm k}=\frac 1{L^4}\sum_{{i},{j}}e^{2\pi i({\bm r}_i-{\bm r}_j)\cdot{\bm k} }(\langle { D^\alpha}_{i} { D^\alpha}_{ j}\rangle-\langle {D^\alpha}_{ i}\rangle \langle{ D^\alpha}_{ j}\rangle) ,
\end{equation}
where $D^x_{ i} = {\bm{s}}_{ i} \cdot {\bm{s}}_{{ i}+(1,0)}$ and $D^y_{ i} = {\bm{s}}_{ i} \cdot {\bm{s}}_{{ i}+(0,1)}$. The dimer order parameter $D^x_{\bm{k}}$ at ${\bm{k}} = (\pi, \pi)/L$ is plotted in Fig.~\ref{fig:order_para}(b), and quadratic finite-size scaling is performed to obtain the results in the thermodynamic limit ($L \rightarrow \infty$). In the thermodynamic limit, the dimer order parameters are $0$ for both cases of $J_2 = 0$ and $J_2 = 0.5 J_1$, which is consistent with reports that the model has no dimer order for both cases \cite{liu2022gapless}.

There are a few remarks that we would like to address. First, in this work, we have only considered translational symmetry. We may further utilize the crystal symmetry, such as the $D_4$ symmetry of the square lattice, to achieve a more accurate result. Second, although we only discuss the spin model in this work, our method is also suited for bosonic or fermionic models.  Furthermore, this approach can be easily generalized to cylindrical or even twisted boundary conditions\cite{niu1985,Supplementary}.

{\it Summary:} We propose a tensor network ansatz named sPEPS for quantum many-body systems with translational symmetry under PBC. We benchmark this method against the Heisenberg model and the $J_1$-$J_2$ model, demonstrating its capability to yield highly accurate results at low computational costs, even for large system sizes. This work marks a significant advancement in PEPS by effectively addressing the challenges posed by PBC. The sPEPS approach can be easily adapted to various boundary conditions, such as cylindrical and twisted boundaries, thus paving the way for broader and more impactful applications in quantum many-body physics. This breakthrough enhances the potential of PEPS methods, providing new tools and perspectives for exploring complex quantum systems.

This work was funded by the Chinese National Science Foundation Grant Numbers 12134012, 12304552, 12104433 and the Strategic Priority Research Program of Chinese Academy of Sciences Grant Number XDB0500201, the Innovation Program for Quantum Science and Technology Grant Number 2021ZD0301200. The numerical calculations in this study were carried out on the ORISE Supercomputer and the USTC HPC facilities.

%

\end{document}